\documentclass[runningheads]{llncs}
\usepackage[T1]{fontenc}
\usepackage{graphicx}
\usepackage{balance} 
\usepackage{float}
\usepackage{algorithm}
\usepackage{algpseudocode}
\usepackage{wrapfig}
\usepackage{xcolor}
\usepackage{svg}
\begin{document}
\title{Frequent Itemset Mining using QUBO}
%
%
\author{Jonas Nüßlein}
%
%
\institute{LMU Munich
\email{jonas.nuesslein@ifi.lmu.de}}
\maketitle              
\begin{abstract}
In this paper we propose a R-step approximation to solve frequent itemset mining on quantum hardware like quantum annealing or QAOA. The idea is to search for the set of items where the minimal 2-item frequency is maximal. This can be represented as a maximum clique problem.

\end{abstract}
\section{Introduction}
Frequent Itemset Mining (FIM) is a common problem in data mining \cite{fim1}. The goal of FIM is to find a set of objects that frequently occur together. An example is the analysis of customer behavior \cite{jonas}: let a set of transactions be given, where a transaction is a multi-set of objects (the objects bought by one person). The goal is then to find a set of objects of size $k$, which were frequently bought together.
\par
It has been shown that FIM is NP-hard \cite{fimnphard} and therefore difficult to solve on classical computers. In this paper we present a simple method of formulating FIM as a Quadratic Unconstrained Binary Optimization (QUBO) \cite{qa2} which can be solved on quantum hardware for example using Quantum Annealing or QAOA \cite{qaoa}.

\section{Background}
\subsection{Quadratic Unconstrained Binary Optimization}
Given a symmetric (n$\times$n)-matrix $Q$ and a binary vector $x$ of length $n$, a QUBO \cite{qa1} \cite{qa3} is a function of the form:

\begin{equation}
    H(x,Q)=\sum_{i=1}^{n}\sum_{j=i}^{n}{x_i\ x_j\ Q_{ij}}
\end{equation}

The optimization task is to find a binary vector $x$ which is as close to the optimum $x^{\ast}=argmin_x \; H(x,Q)$ as possible. This we want to delegate to the machine. Our task, on the other hand, is to specify a function which maps a FIM problem instance $P$ (a database of transactions) to a QUBO matrix $Q$ in such a way that the solution $p$ (i.e. the frequent itemsets) for the problem instance $P$ can be derived from the solution vector $x^{\ast}$.

\subsection{Maximum Clique}
Maximum clique is the problem of finding the biggest set $C \subseteq V$ of vertices in a graph $G = (V,E)$ such that all vertices in $C$ are pairwise connected. This problem is known to be NP-complete \cite{karp21}.

\section{FIM as QUBO}
We are looking for a frequent itemset of size $N$. The general idea is as follows: we model the database as a graph, where the vertices represent the objects and the edge weights $w_{ij}$ are the empirical probabilities that the two connected vertices $i$ and $j$ are in one transaction. We then conduct a R-step approximation: we set the threshold to $\tau = 0.5$ then we remove all edges from a graph $G$ with edge weight lower than $\tau$. For the resulting graph (which is now not necessarily fully connected) we calculate the maximum clique. The clique does not necessarily has size $N$. The next step is to adjust the threshold $\tau$ and repeat the process. If the size of the found maximum clique is lower then $N$ then we reduce $\tau$, if the clique size was larger then $N$ then we increase $\tau$. So the optimization problem we are solving can be formulated as:

\begin{equation}
\max\limits_{I} \min\limits_{i,j \in I} P(i \cup j)
\end{equation}

$I$ is the frequent itemset we are searching and $P(i \cup j)$ is the empirical probability of objects $i$ and $j$ occuring in the same transaction. The whole algorithm is presented in Algorithm 1.

\begin{algorithm}[t]
\caption{FIM as QUBO}\label{alg:cap}
\begin{algorithmic}[1]
\Require
\State $F_{ij}$ \Comment{Empirical probability of items $i$ and $j$ occuring in the same transaction }
\State $K$ \Comment{Number of objects}
\State $\tau = 0.5$ \Comment{Threshold}
\State $solution = []$
\State $N$
\State $R$ \Comment{number of iterations}

\vspace{0.2cm}

\For{$i = 1$ to $R$}
\State $Q = [][]$
\For{$q_1 = 0$ to $K-1$}
\For{$q_2 = 0$ to $K-1$}
\If{$q_1 == q_2$}
\State $Q[q_1][q_2] = -1$
\ElsIf{$F_{q_1q_2} < \tau$}
\State $Q[q_1][q_2] = K+1$
\EndIf
\EndFor
\EndFor

\vspace{0.2cm}

\State $answer$ = solve($Q$) \Comment{Solve QUBO with quantum hardware}

\vspace{0.2cm}

\If{size of clique in answer $\geq$ $N$}
\State $solution$ = getSelectedVertices($answer$)
\State $\tau = \tau + 2^{-(i+1)}$
\Else
\State $\tau = \tau - 2^{-(i+1)}$
\EndIf

\EndFor

\end{algorithmic}
\end{algorithm}

\section{Experiments}
As a little proof of concept experiment we created 24000 completely random transactions of size 22 out of a basic quantity of 250 items. Additionally, we added the following transaction 1000-times to the database [1,2,3,…,20]. So, in total the database consisted of 25,000 transactions. We then computed the most frequent 20-item set (which was clearly [1,2,3,…,20]).
\par
MFIO returned the correct solution in 34 seconds. For comparison we computed the most frequent 20-set also with Bomo. Bomo is based on FPGrowth \cite{fpgrowth}, but it only searches for the top-N (in our case the top-1) k-itemsets. So, it doesn’t need a threshold \cite{asdf}. We used a fast C implementation of Bomo \cite{bomo}. Bomo returned the correct result after 102s. Also, the memory usage of MFIO was significantly more efficient (measured was the memory needed for the internal data structures):
\begin{itemize}
    \item MFIO: 122KB (for the 31.125 2-sets)
    \item Bomo: 83MB
\end{itemize}
\ \\
As a little real-world example we were able to find the most frequent 4-, 5-, 6- and 7-set letters in english words. So, the basic quantity was the alphabet and the words (subsets of the alphabet) were the transactions. The database consisted of 370.000 english words \cite{englishwords}. The computed sets were:
$\{a,e,i,n\}$,
$\{a,e,i,n,t\}$,
$\{a,e,i,n,s,t\}$,
$\{a,e,i,n,o,s,t\}$
All experiments were conducted on D-Wave 2000Q.

\section{Related Work}
FP-Growth \cite{fpgrowth} is the most common approach for frequent itemset mining. Bomo \cite{bomo} is a variant of FPGrowth which only searches for k-itemsets. The most related work to this paper is \cite{maxclique}. The authors already propose to use maximum clique to find frequent itemsets. However they use a user-defined threshold. We extend this work by proposing a R-step approximation and solving the maximum clique problems on quantum hardware.
\par
In \cite{quantumassociation} a quantum algorithm for finding association rules is presented. In contrast to that we are searching for the most frequent itemset of length $N$ with QUBO.
\par
In \cite{maxclique1} and \cite{maxclique2} QUBO and Ising formulations for maximum clique are presented. Our approach builds on these formulations.

%
%
%
%

\end{document}